\documentclass[times]{qjrms3}

\usepackage{moreverb}

\usepackage{amsmath}
\usepackage{amssymb}

\usepackage{natbib}
\DeclareMathOperator*{\sumsum}{\sum\sum}

\def\volumeyear{2008}

\begin{document}

\pagenumbering{gobble}
\twocolumn[
This is the peer reviewed version of the following article: Shima, S., Kusano, K., Kawano, A.,
Sugiyama, T. and Kawahara, S. (2009), The super-droplet method for the numerical simulation of
clouds and precipitation: a particle-based and probabilistic microphysics model coupled with a nonhydrostatic
model. Q.J.R. Meteorol. Soc., 135: 1307-1320, which has been published in final form at
https://doi.org/10.1002/qj.441. This article may be used for non-commercial purposes in accordance
with Wiley Terms and Conditions for Use of Self-Archived Versions. This article may not be enhanced,
enriched or otherwise transformed into a derivative work, without express permission from Wiley or
by statutory rights under applicable legislation. Copyright notices must not be removed, obscured or
modified. The article must be linked to Wiley’s version of record on Wiley Online Library and any
embedding, framing or otherwise making available the article or pages thereof by third parties from
platforms, services and websites other than Wiley Online Library must be prohibited.
]
\clearpage
\setcounter{page}{1}
\pagenumbering{arabic}

\runningheads{S.~Shima et al.}{
Super-Droplet Method for the Numerical Simulation of Clouds and Precipitation:
}

\title{
Super-Droplet Method for the Numerical Simulation of Clouds and Precipitation: a Particle-Based and Probabilistic Microphysics Model Coupled with Non-hydrostatic Model
}

\author{
S.~Shima,\corrauth\ K.~Kusano,\ A.~Kawano,\ T.~Sugiyama\ and S.~Kawahara
}

\address{
The Earth Simulator Center, 
Japan Agency for Marine-Earth Science and Technology, 
Yokohama, 
Japan
}

\corraddr{
S.~Shima,
The Earth Simulator Center,
Japan Agency for Marine-Earth Science and Technology,
3173-25 Showa-machi, 
Kanazawa-ku, 
Yokohama City, 
Kanagawa, 
236-0001, 
JAPAN.
E-mail: s\_shima@jamstec.go.jp
}

\begin{abstract}
A novel, particle based, probabilistic approach for the simulation of cloud
microphysics is proposed, which is named the Super-Droplet Method
(SDM). This method enables accurate simulation of cloud microphysics with less
demanding cost in computation.  SDM is applied to a warm-cloud system, which
incorporates sedimentation, condensation/evaporation, and stochastic
coalescence.  The methodology to couple super-droplets
and a non-hydrostatic model is also developed.  It is confirmed that
the result of our Monte Carlo scheme for the stochastic coalescence of
super-droplets agrees fairly well with the solutions of the stochastic
coalescence equation.  
The behavior of the model is evaluated using a simple test problem,
that of a shallow maritime
cumulus formation initiated by a warm bubble.
Possible extensions of SDM are briefly discussed.
A theoretical analysis suggests that the computational cost of SDM
becomes lower than the spectral (bin) method when the number of
attributes - the variables that identify the state of each
super-droplet - becomes larger than some critical value, which we estimate to be in the range 
$2\sim4$.
\end{abstract}


\received{xx April 2008}
\revised{\quad}
\accepted{\quad}

\maketitle

%\footnotetext[2]{Please ensure that you use the most up to date
%class file,
%available from the QJRMS Home Page at\\
%\href{http://www.interscience.wiley.com/qj}{\texttt{http://www.interscience.wiley.com/qj}}%
%}

\section{Introduction}
\label{1}
Although clouds play a crucial role in atmospheric phenomena, the
numerical modeling of clouds remains somewhat crude.
Clouds incorporate both cloud dynamical processes (i.e., turbulent
fluid dynamics), and cloud microphysical processes (such as kinetic
interactions among particles, sedimentation, etc.)
These two processes mutually affect each other in the
course of cloud and precipitation development, and this
fact suggests that we need to simulate both processes and their
interactions concurrently in order to produce an accurate prediction
\citep[e.g.,][]{Stevens1998}.
Cloud dynamics model to describe the fluid motion in the atmosphere
has been well developed
and have been widely used since the 1960s
\citep[e.g.,][]{Ogura1962,Lilly1962,Clark1973}.
However, it
is still difficult to perform an accurate simulation of cloud
microphysical processes, though several simulation methods
have been proposed.
For example, the bulk
parameterization method 
\citep[e.g.,][]{Kessler1969,Ziegler1985,Murakami1990,Ferrier1994,Meyers1997,Khairoutdinov2000, Seifert2001}
is computationally inexpensive but less accurate, 
where as the spectral (bin) method 
\citep[e.g.,][]{Clark1973,Soong1974,Hall1980,Kogan1991,Stevens1996,Khain2004_a}
can be accurate but computationally demanding. 
Accurate and inexpensive numerical methods to simulate the cloud microphysics and
the interactions between the cloud dynamics are required to understand
and predict cloud-related phenomena.

In this paper we propose a novel, particle based, probabilistic
approach for the simulation of cloud microphysics, named the
Super-Droplet Method (SDM).  As an illustrative example, SDM is
applied to a warm-cloud system, which incorporates
sedimentation, condensation/evaporation, and stochastic coalescence.
The methodology to couple super-droplets and a non-hydrostatic model
is also developed. 
The extensions and the computational efficiency of the SDM
are also briefly discussed. It is suggested that SDM is as accurate as the
spectral (bin) method, but could be less demanding in computation
when we simulate many detailed cloud microphysical processes. 
The most expensive part of the spectral (bin) method is the
stochastic coalescence process because it appears in the form of a
multiple-integral term. 
However, the particle based and probabilistic approach of SDM
enables us to reduce the computational cost.  

The organization of the present paper is as the following. In Sec.~\ref{2}
we introduce {\it the primitive model}, which is a detailed
microphysics-dynamics coupled warm-cloud model. In this paper this
model is considered as the master framework for describing the
behavior of warm clouds, and will be the starting point for our
discussions. In Sec.~\ref{3} we review the traditional methods to simulate cloud microphysics,
and see how they can be applied to the primitive model.
In Sec.~\ref{4} and \ref{5}, we apply SDM to the primitive model
in two steps.  In Sec.~\ref{4}, introducing the notion of {\it super-droplet},
we derive a coarse-grained model of the
primitive model, and discuss the basic ideas of SDM. The model introduced in this section cannot be
implemented on a computer directly, and a numerical simulation scheme
is proposed for each process in Sec.~\ref{5}.  In particular, a novel Monte
Carlo scheme for the stochastic coalescence process is developed and
validated. In Sec.~\ref{6}, the behavior of the models is evaluated using a simple test problem, that of a
shallow maritime
cumulus formation initiated by a warm bubble.
This is introduced as a proof of concept of our ideas, and more quantitative
and detailed evaluations of the methods will be pursued in subsequent
work.
In Sec.~\ref{7} possible extensions of the methods we develop are briefly discussed 
as are some of the computational issues involved.
Finally, a
summary and concluding remarks are presented in Sec.~\ref{8}.

%%%%%%%%%%%%%%%%%%%%%%%%%%%%%%%%%%%%%%%%%%%%%%%
\section{The primitive model}
\label{2}
This section is devoted for the introduction of {\it the primitive
model}. This is a detailed microphysics-dynamics coupled warm-cloud
model in the sense that it follows all the aerosol/cloud/precipitation
particles in the atmosphere.  Though there are various cloud
microphysical processes \citep[e.g.,][]{Pruppacher}, we selected three
elementary processes which could be the minimum for describing
warm clouds.  In the proceeding sections, we will see how 
the traditional methods and
the SDM can simulate the primitive model.

\subsection{Cloud microphysics of the primitive model}
\label{2.1}
\subsubsection{Definition of droplet}
\label{2.1.1}
We use the word {\it droplet} as a generic term referring  
to the aerosol/cloud/precipitation particles.

Let $N_r(t)$ be the number of droplets in some specified atmospheric control volume 
at time $t$. The $i$-th droplet is completely characterized by a
set of variables $\{\mathbf x_i(t), \mathbf v_i(t), R_i(t), M_i(t)\}$.  Here,
$i=1,2,\dots,N_r(t)$; $\mathbf x_i(t)$ is the position coordinate of the
droplet; $\mathbf v_i(t)$ is its velocity;
$R_i(t)$ is the equivalent radius, which represents the amount of
water that the droplet contains, defined as the radius of a
sphere having the same volume as the contained water; $M_i(t)$ denotes
the mass of the solute contained in the droplet.  In general,
several sorts of soluble/insoluble aerosols are contained in each
droplets.  However, in this model, we consider aerosols are
composed of only one soluble substance (solute) for simplicity.

We refer to the following state variables as the attributes of
droplets: the velocity $\mathbf v_i(t)$; the radius $R_i(t)$; and the
mass of solute $M_i(t)$.  To simplify the notation, the attributes are
sometimes denoted by $\mathbf a_i(t)$.  Note that we do not regard the
position $\mathbf x_i(t)$ as an attribute for convenience. 

As we will see in the next section, we assume that the droplet
velocity immediately reaches their terminal velocity, and hence $\mathbf
v_i(t)$ is determined by their radius $R_i(t)$. As a result, the
number of independent attributes of our droplet is two: the
radius $R_i(t)$ and the mass of solute $M_i(t)$. 

\subsubsection{Motion of droplets}
\label{2.1.2}
Assuming that all the droplets reach their terminal velocity
immediately, the motion of each droplet is governed by the following equations:
\begin{equation}
\label{eq_motion}
\mathbf v_i(t)=\mathbf
U(\mathbf x_i)-\hat{\mathbf z}v_{\infty}(R_i),\quad \frac{d\mathbf x_i}{dt} =\mathbf v_i,
\end{equation}
where $\mathbf U(\mathbf x_i)$ is the wind velocity at the droplet position $\mathbf
x_i(t)$ and $v_{\infty}(R_i)$ is given by the
semi-empirical formulas developed by \cite{Beard1976} \citep[see also][Chap.~10]{Pruppacher}.
Consequently, $\mathbf v_i(t)$ is no longer an
independent attribute of droplet in our model.

\subsubsection{Condensation/evaporation of droplets}
\label{2.1.3}
We consider the mass change of droplets through the
condensation/evaporation process according to K\"{o}heler's theory,
which takes into account the solution and curvature effects on the
droplet's equilibrium vapor pressure \citep[][Chap.~13]{Kohler1936, Rogers, Pruppacher}.
The growth equation of the radius $R_i$ is derived as the following.
\begin{gather}
\label{eq_growth}
 R_i \frac{dR_i}{dt} = \frac{(S-1) - \frac{a}{R_i} + \frac{b}{R_i^3}}{F_k + F_d},\\
F_k =\left(\frac{L}{R_v T} -1\right)\frac{L\rho_\mathrm{liq}}{KT},\quad
F_d =\frac{\rho_\mathrm{liq} R_v T}{D e_s(T)} .\notag
\end{gather}
Here, $S$ is the ambient saturation ratio; $F_k$ represents the
thermodynamic term associated with heat conduction; $F_d$ is the term
associated with vapor diffusion; the term $a/R_i$ represents the
curvature effect which expresses the increase in saturation ratio over
a droplet as compared to a plane surface; the term $b/R_i^3$ shows the
reduction in vapor pressure due to the presence of a dissolved
substance and $b$ depends on the mass of solute $M_i$ dissolved in the
droplet. Numerically, $a \simeq 3.3 \times 10^{-5}~\mathrm{cm~K}/T$
and $ b \simeq 4.3~\mathrm{cm}^3~ i M_i/m_s$, where $T$ is the
temperature, $i\simeq2$ is the degree of ionic dissociation, $m_s$ is
the molecular weight of the solute.
$R_v$ is the individual gas constant for water vapor, $K$ is the
coefficient of thermal conductivity of air, $D$ is the molecular
diffusion coefficient, $L$ is the latent heat of vaporization, 
$e_s(T)$ is the saturation vapor pressure, and 
$\rho_\mathrm{liq}=1.0~\mathrm{g~cm^{-3}}$ is the density of liquid water.

\subsubsection{Stochastic coalescence of droplets}
\label{2.1.4}
Two droplets may collide and coalesce into one bigger droplet
and this process is responsible for precipitation development.  It is
worth noting here that the coalescence process is our particular
concern in this paper because SDM is expected to overcome the difficulty in the
numerical simulation of the coalescence process.  

Our primary postulate, which is commonly used in many previous
studies, is that the coalescence process can be described in a
probabilistic way: Following, e.g., \cite{Gillespie1972}, consider a region with volume $\Delta V$.  If
$\Delta V$ is small enough, we can consider that the droplets
inside this region are ``well-mixed,'' and the droplets coalesce
with each other with some probability during a sufficiently short time
interval $(t,t+\Delta t)$, i.e., there exists such $C(R_j, R_k)$ that
satisfies
\begin{equation}
\label{prob}
\begin{split}
P_{jk} 
=& C(R_j, R_k)
 | \mathbf v_j - \mathbf v_k| \frac{\Delta t}{\Delta V}\\ 
=& K(R_j, R_k)\frac{\Delta t}{\Delta V}\\ 
=&\mbox{probability that droplet $j$ and $k$ } \\ 
 &\mbox{inside a small region $\Delta V$ will coalesce }\\
 &\mbox{in a short time interval $(t,t+\Delta t)$.}
\end{split}
\end{equation}
Here, $C(R_j, R_k)$ can be regarded as the effective collision
cross-section of the droplets $j$ and $k$ and may be evaluated as
\begin{equation}
\label{col_eff}
C(R_j, R_k)=E(R_j, R_k)\pi (R_j+R_k)^2,
\end{equation}
where $E(R_j, R_k)$ is the collection efficiency, which takes into
account the effect that a smaller droplet would be swept aside by the
stream flow around the larger droplet or bounce on the surface 
\citep[][Chap.~14]{Davis1972,Hall1980,Jonas1972,Pruppacher}. $K(R_j, R_k)$ in
\eqref{prob} is defined by $K(R_j, R_k):=C(R_j, R_k)| \mathbf v_j - \mathbf v_k|$
and it is called the coalescence kernel, which will later appear in
the stochastic coalescence equation.

Here, let us estimate the typical size of the well-mixed volume
$\Delta V$ by a dimensional analysis.  We can consider that the
droplets inside the clouds are being mixed by the turbulence. Let
$\epsilon$ be the energy dissipation rate of the cloud turbulence, $t$
be the typical time scale of the evolution of the droplet size
distribution, and $l$ be the typical length scale of the well-mixed
volume $\Delta V$. Comparing the dimensions of these three variables,
we get $l=\epsilon^{1/2}t^{3/2}$. Typical value of $\epsilon$ is
$10^{-3}~\mathrm{m^2~s^{-3}} \sim 10^{-1}~\mathrm{m^2~s^{-3}}$
\citep{Siebert}, and $t$ is $100~\mathrm{s}$ (which will be justified
in Fig.~\ref{fig.golovin}). Hence, $l$ ranges from $30~\mathrm{m}$ to
$300~\mathrm{m}$ depending on the intensity of the turbulence, and we
can say that the well-mixed assumption can be justified if $\Delta V$
is smaller than $l^3$.

The cloud microphysical process of the primitive model has been defined.
We will turn to define the cloud dynamical process of the primitive model.

\subsection{Cloud dynamics of the primitive model}
\label{2.2}
We adopt a non-hydrostatic model to describe the cloud dynamical
process of the primitive model.  There exists several variations of
non-hydrostatic model, but the basic equations used in this paper is
as the following.
\begin{gather}
\rho \frac{D\mathbf U}{Dt}=-\nabla P -(\rho+\rho_w)\mathbf g + \lambda \nabla^2\mathbf U,\label{motion}\\ 
P=\rho R_d T,\label{sate}\\
\frac{D\theta}{Dt}=-\frac{L}{c_p\Pi}S_v+\kappa\nabla^2\theta,\label{adiabatic}\\
\frac{D\rho}{Dt}=-\rho\nabla\cdot\mathbf U,\label{continuity}\\ 
\frac{Dq_v}{Dt}=S_v+\kappa\nabla^2q_v.\label{qv}
\end{gather}
Here, $D/Dt:=\partial/\partial t +\mathbf U \cdot \nabla$ is the material
derivative; $\rho=\rho_d+\rho_v$ is the density of moist air, which is
represented by the sum of dry air density $\rho_d$ and vapor density
$\rho_v$; $q_v=\rho_v/\rho$ is the mixing ratio of vapor; $\mathbf U$ is
the wind velocity; $T$ is the temperature; $\theta$ is the potential
temperature; $\Pi=(P/P_0)^{(R_d/c_p)}$ is the Exner function with
reference pressure $P_0$; $\rho_w$ is the density of liquid water per unit volume of air;
$S_v$ is the source term of water vapor associated with
condensation/evaporation process; $\mathbf g$ is the gravitational
constant; $\lambda$ and $\kappa$ are the transport coefficient; $R_d$ is
the gas constant for dry air; $c_p$ is the specific heat of dry air at
constant pressure; $L$ is the latent heat of vaporization.  Equations
\eqref{motion}-\eqref{qv} represent equation of motion, equation of
state, thermodynamic equation, mass continuity, and water continuity,
respectively. Note here that it is assumed that $q_v\ll1$ to derive these equations.  

There are three coupling terms from the microphysics: 
$\rho_w$ is the density of liquid water per unit volume of air, 
which represents the momentum coupling; 
$S_v$ is the source term of vapor through the condensation/evaporation
process; 
$LS_v/c_p\Pi$ is the release of latent heat through the
condensation/evaporation process.  
$\rho_w$ and $S_v$ can be evaluated by the microphysics variables as,
\begin{gather}
\label{eq.10}
\rho_w(\mathbf x,t):=\sum_{i=1}^{N_r}m_i(t)\delta^3(\mathbf x-\mathbf x_i(t)),\\
\label{eq.11}
S_v(\mathbf x,t):=\frac{-1}{\rho(\mathbf x,t)}\sum_{i=1}^{N_r}\frac{dm_i(t)}{dt}\delta^3(\mathbf x-\mathbf x_i(t)),
\end{gather}
where $m_i:=(4\pi/3)R_i^3\rho_\mathrm{liq}$ is the mass of the droplet
$i$. 

The primitive model has now been defined completely and we can see that
this is a detailed microphysics-dynamics coupled warm-cloud
model. 

%%%%%%%%%%%%%%%%%%%%%%%%%%%%%%%%%%%%%%%%%%%%%%%
\section{Traditional methods for the simulation of cloud microphysics}
\label{3}
In this section, we review the traditional methods to simulate the
cloud microphysics and see how they can simulate the primitive
model and what are the problems involved in each method.

\subsection{Direct simulation using the exact Monte Carlo method}
\label{3.1}
If we can follow all the the droplets, we can perform a direct 
simulation of the cloud microphysical processes.  To do that, we
have to store all the information of the droplets, $\{(\mathbf x_i(t),
\mathbf a_i(t)) \mid i=1,2,\dots,N_r(t)\}$, on the memory of our computer.  The
motion and condensation/evaporation of droplets can be simulated
by solving the ordinary differential equations \eqref{eq_motion} and
\eqref{eq_growth} for all the droplets.  A direct simulation of
the stochastic coalescence process \eqref{prob} can also be performed
if we use the exact Monte Carlo method, which was developed by
\cite{Gillespie1975} and improved by 
\cite{Seeszelberg1996}. Their procedure repeatedly
draws a random waiting time for which the next one pair of
droplets will coalesce.

Therefore it is conceptually possible to perform an exact simulation
of the cloud microphysics in a sufficiently large region to
simulate the cloud formation and precipitation. However, even the
world most powerful computer can not do that because of the very high
cost in computation: the typical number density of droplets is
$10^7\sim10^9\mathrm{m^{-3}}$.

\subsection{Bulk parameterization method}
\label{3.2}
The bulk parameterization method is the most common way to deal with
cloud microphysics 
\citep[e.g.,][]{Kessler1969,Ziegler1985,Murakami1990,Ferrier1994,Meyers1997,Khairoutdinov2000,Seifert2001}.
The characteristic feature of this method is that the cloud
microphysical processes are approximately represented by the dynamics
closed in a very small number of variables, such as cloud water mass
content, rain water mass content, and their number concentrations.
Consequently, the number of degrees of freedom is reduced so much that
the computational demand is extremely relaxed.  However, 
to derive the equations from the primitive model,
we need to rely on several empirical assumptions, 
such as, on the function form of the droplet size distribution.
Hence, this methodology should further be elaborated for accurate simulation 
\citep[e.g.,][]{Seifert2001,Khain2004_b,Lynn2005_a,Lynn2005_b}.

% Even worse, there is no guarantee whether we can reduce the cloud microphysics process or not

\subsection{Spectral (bin) method}
\label{3.3}
Another approach to the cloud microphysics is the spectral (bin) method
\citep[e.g.,][]{Clark1973,Soong1974,Hall1980,Kogan1991,Stevens1996,Khain2004_a}
 The spectral (bin)
method is a class of numerical schemes which solves the time
evolution equation of the number density distribution of droplets
by discretizing the number density distribution into bins. 

The time evolution equation of the number density distribution is sometimes called 
the stochastic coalescence equation (SCE). 
This is an integro-differential equation and can be derived
under the assumption that we can ignore a certain correlation among
the droplet population probabilities 
\citep{Gillespie1972}. (The explicit form of the SCE is displayed in Appendix~\ref{app.B}~\eqref{eq.SCE_general}.)
This assumption seems to be correct 
\citep{Seeszelberg1996}, thus the result of the spectral (bin)
method could be very accurate. However, the computational cost of
the spectral (bin) method would be very expensive if we consider various
microphysical processes and hence the number of attributes $d$ becomes
large. Some discussions on this point will be carried out in Sec.~\ref{7} and Appendix~\ref{app.B},
but in a word, this is owing to the fact that the SCE contains a
$d$-multiple integral.

%%%%%%%%%%%%%%%%%%%%%%%%%%%%%%%%%%%%%%%%%%%%%%%
% Super-Droplet Method: theoretical modeling
\section{Super-Droplet Method: basic equations}
\label{4}
In the two subsequent sections SDM is applied to the primitive model.
In this section, introducing the notion of {\it super-droplet}, we derive the basic equations of the SDM,
which can be regarded as a coarse-grained model of the primitive model. 
The numerical implementations will be proposed in the next section.

\subsection{Cloud microphysics of Super-Droplet Method}
\label{4.1}
\subsubsection{Definition of super-droplet}
\label{4.1.1}
First of all, let us define what a super-droplet is.  Each
super-droplet represents a multiple number of droplets with the
same attributes and position, and the multiplicity is denoted by the
positive integer $\xi_i(t)$, which  can be
different in each super-droplet, and time-dependent due to the
definition of coalescence introduced later. That is to say, each
super-droplet has its own position $\mathbf x_i(t)$ and its own attributes
$\mathbf a_i(t)$, which characterize the $\xi_i(t)$ identical
droplets represented by the super-droplet $i$.  Here, for our
simple warm-rain system, the attribute consists of the equivalent
radius of water and the solute mass: $\mathbf a_i(t) = (R_i(t), M_i(t)).$
Note here that no two droplets have exactly the same position and
attributes, and in this sense, super-droplet is a kind of coarse-grained
view of droplets both in real-space and attribute-space.  Let
$N_s(t)$ be the number of super-droplets floating in the atmosphere at
time $t$.  Then, these super-droplets represent
$N_r(t)=\sum_{i=1}^{N_s(t)}\xi_i(t)$ number of droplets in total.
To summarize, we approximate the droplets $\{(\mathbf x_i(t), \mathbf a_i(t)) \mid
i=1,2,\dots,N_r(t)\}$ by the super-droplets $\{(\mathbf
x_j(t), \mathbf a_j(t), \xi_j(t)) \mid j=1,2,\dots,N_s(t)\}$.  In the
following, we give the time evolution law of the super-droplets.

\subsubsection{Motion of super-droplets}
\label{4.1.2}
Except for the coalescence process, each super-droplet behaves in just the
same way as a droplet. Thus, a super-droplet moves according to
the motion equation \eqref{eq_motion} and $\mathbf v_i(t)$ is evaluated by
the terminal velocity.

\subsubsection{Condensation/evaporation of super-droplets}
\label{4.1.3}
The condensation/evaporation process of super-droplets is governed by
the growth equation \eqref{eq_growth}.

\subsubsection{Stochastic coalescence of super-droplets}
\label{4.1.4}
The stochastic coalescence process for the super-droplets must be
formulated carefully so that the super-droplets well-approximate the
behavior of droplets.  The formulation procedure is in two steps:
1) define how a pair of super-droplets coalesce; 2) determine the
probability that the super-droplet coalescence occurs.

\begin{figure}
\centering
\includegraphics[width=8.5cm]{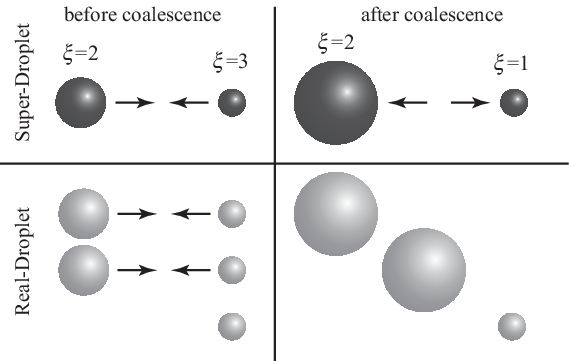}
\caption{Schematic view of the coalescence of super-droplets.  Two
super-droplets with multiplicity $2$ and $3$ undergo coalescence
(upper left). This represents the coalescence of two droplet pairs
(lower left and right). As a result the super-droplet with multiplicity $2$ becomes larger and
the multiplicity of the other super-droplet decreases $3\to1$ (upper right).
\label{fig.1}
}
\end{figure}

Consider the case when a pair of super-droplets $(j,k)$ will
coalesce. These super-droplets represent $\xi_j$ 
droplets $(\mathbf a_j, \mathbf x_j)$ and $\xi_k$ droplets
$(\mathbf a_k, \mathbf x_k)$. Let us define that exactly $\min{(\xi_j,\xi_k)}$
pairs of droplets will contribute to the coalescence of
the super-droplet pair $(j,k)$. Figure.~\ref{fig.1} is a schematic view of
the coalescence of super-droplets.  In this example, $\xi_j=3$ and
$\xi_k=2$. We can see that $\min{(\xi_j,\xi_k)}=2$ pairs of
droplets undergo coalescence, which results in the decrease of
multiplicity $\xi_j:3\to1$ and the increase of the size of the
super-droplet $k$.

Below is the complete definition of how a pair of super-droplets
$(j,k)$ change their state after they coalesce:
\begin{enumerate}
\item if $\xi_j\neq\xi_k$, we
can choose $\xi_j>\xi_k$ without losing generality, and
\begin{gather}
\xi'_j=\xi_j-\xi_k,\quad\xi'_k=\xi_k,\label{eq.15}\\
R'_j=R_j,\quad R'_k=(R_j^3+ R_k^3)^{1/3},\label{eq.16}\\
M'_j=M_j,\quad  M'_k=(M_j+M_k),\label{eq.17}\\
\mathbf x'_j=\mathbf x_j,\quad\mathbf x'_k=\mathbf x_k,\label{eq.18}
\end{gather}
where the dashed valuables represent the updated value after the coalesce.
\item if $\xi_j=\xi_k$, 
\begin{gather}
\xi'_j=[\xi_j/2],\quad \xi'_k=\xi_j-[\xi_j/2],\label{eq.19}\\
R'_j=R'_k=\left(R_j^3+ R_k^3\right)^{1/3},\label{eq.20}\\
M'_j=M'_k=\left( M_j+ M_k\right),\label{eq.21}\\
\mathbf x'_j=\mathbf x_j,\quad\mathbf x'_k=\mathbf x_k,\label{eq.22}
\end{gather}
where Gauss' symbol $[b]$ is the greatest integer that is less than or
equal to the real number $b$. 
\end{enumerate}
Note here that in the case $\xi_j=\xi_k$, no
droplets will be left after the coalescence because all $\xi_j$
number of droplets contribute to the coalescence.  Thus, we
divide the resulting $\xi_j$ number of droplets into two super-droplets.

This definition of super-droplet coalescence possesses a favorable
property that the number of super-droplets is unchanged in most cases
though the number of droplets always decreases.  The number of
super-droplets is decreased through the coalescence only when
$\xi_j=\xi_k=1$, i.e., both super-droplets are droplets.  This
results in $\xi'_j=0$ and $\xi'_k=1$, and we remove the super-droplet
$j$ out of the system.  Because the number of super-droplets
corresponds to the accuracy of SDM, the number conservation of
super-droplets suggests the flexible response of SDM to the drastic
change of the number of droplets.

Having defined how a pair of super-droplets coalesces, we turn to
determine the probability that the coalescence occurs. If we require
the consistency of the expectation value, the probability is determined as
the following.
\begin{equation}
\label{prob_super}
\begin{split}
P_{jk}^{(s)}
=&\max(\xi_j,\xi_k)P_{jk}\\
=&\mbox{probability that super-droplet $j$ and $k$ } \\ 
 &\mbox{inside a small region $\Delta V$ will coalesce }\\
 &\mbox{in a short time interval $(t,t+\Delta t_c)$.}
\end{split}
\end{equation}
Indeed, we can confirm that the expectation value is identical to that
of the primitive model (the real world). The super-droplet $j$
represents $\xi_j$ droplets with attribute $\mathbf a_j$
and the super-droplet $k$ represents $\xi_k$ droplets with
attribute $\mathbf a_k$.  In terms of the real world, this
corresponds to the situation that there are $\xi_j\xi_k$ 
droplet pairs which have the possibility to coalesce with 
probability $P_{jk}$. Thus, the number of droplet pairs which
will coalescence follows the binomial distribution with $\xi_j\xi_k$
trials and success probability $P_{jk}$. Thus, in the real
world, the expectation value of the number of coalesced pairs is
$\xi_j\xi_kP_{jk}$. On the other hand, in the super-droplet world, a
coalescence of the super-droplets $j$ and $k$ represents the
coalescence of $\min(\xi_j,\xi_k)$ pairs of droplets with
attribute $\mathbf a_j$ and $\mathbf a_k$.  Thus, the coalescence of
the super-droplets $j$ and $k$ is expected to represent
$\min(\xi_j,\xi_k)P_{jk}^{(s)}=\min(\xi_j,\xi_k)\max(\xi_j,\xi_k)P_{jk}=\xi_j\xi_kP_{jk}$
coalescences of droplet pairs, which is identical to the
value of the real world. It is worth noticing here that the
variance in the super-droplet world becomes larger than that of the
real world. In the real world, the variance of the
number of coalesced pairs is $\xi_j\xi_kP_{jk}(1-P_{jk})\simeq
\xi_j\xi_kP_{jk}=:V_r$ (Poisson distribution limit).  In the
super-droplet world, the variance is $\{\min(\xi_j,\xi_k)\}^2
P_{jk}^{(s)} - \{ \xi_j\xi_kP_{jk}\}^2\simeq \min(\xi_j,\xi_k) V_r$,
which is $\min(\xi_j,\xi_k)$ times larger than that of the
real world.

We have defined the cloud microphysics of super-droplets.  If
all the multiplicity $\xi_i=1$ then our model is equivalent to the
primitive model. Thus, we can change the accuracy of our model by
changing the initial distribution of multiplicity $\{\xi_i(t=0)\}$, or
equivalently, the initial number of super-droplets $N_s(t=0)$. 
If the convergence to the primitive model is rapid, we can say that
the use of SDM is beneficial.
%We are
%expecting the rapid convergence to the primitive model so that the
%super-droplet method to be useful.

The definition of the coalescence process of super-droplets is still
theoretical and we have to do numerical simulation on this model
somehow.  In Sec.~\ref{5.1.3}, we will develop a Monte Carlo scheme for the
stochastic coalescence, which can simulate the coalescence process of
super-droplets efficiently.

\subsection{Cloud dynamics of Super-Droplet Method}
\label{4.2}
For the cloud dynamics of SDM, we can also apply the non-hydrostatic
model \eqref{motion}-\eqref{qv}. Note here that we have to evaluate
the source terms from microphysics \eqref{eq.10} and \eqref{eq.11} not by
the droplets but by the super-droplets:
\begin{gather}
\label{eq.src_super1}
\rho_w(\mathbf x,t)=\sum_{i=1}^{N_s}\xi_im_i(t)\delta^3(\mathbf x-\mathbf x_i(t)),\\
\label{eq.src_super2}
S_v(\mathbf x,t):=\frac{-1}{\rho(\mathbf x,t)}\sum_{i=1}^{N_s}\xi_i\frac{dm_i(t)}{dt}\delta^3(\mathbf x-\mathbf x_i(t)).
\end{gather}

%%%%%%%%%%%%%%%%%%%%%%%%%%%%%%%%%%%%%%%%%%%%%%%
% Super-Droplet Method: numerical implementation}
\section{Super-Droplet Method: numerical implementation}
\label{5}
The model developed in the previous section is still theoretical.  In
this section, we propose a numerical simulation scheme, in which we
solve the individual processes independently in  different time steps,
instead of solving the whole equations at once.  In particular, a novel
Monte Carlo scheme for the stochastic coalescence process is developed
and validated.

\subsection{How to simulate cloud microphysics}
\label{5.1}
\subsubsection{Motion of super-droplets}
\label{5.1.1}
Let $\Delta t_m$
be the time step to simulate this process.  After evaluating the
terminal velocity $\mathbf v_i(t)=\mathbf U_i^* -\hat{\mathbf z}v_{\infty}(R_i(t))$, 
we can update the position as
$\mathbf x_i(t+\Delta t_m)= \mathbf x_i(t) + \Delta t_m\mathbf
v_i(t)$. Here, $\mathbf U_i^*$ is the wind velocity at the
super-droplet position $\mathbf x_i(t)$. As we will see in
Sec.~\ref{5.2}, we use a regular grid $\mathbf x_{lmn}$ for the
spatial discretization of  the fluid field variables. Let us evaluate
$\mathbf U_i^*$ by the linear interpolation of the current
wind-velocity field $\mathbf U(\mathbf x_{lmn},t)$.

\subsubsection{Condensation/evaporation of super-droplets}
\label{5.1.2}
Let $\Delta t_g$ be the time step to simulate this process.  Because
the time scale of condensation/evaporation can be much faster than
$\Delta t_g$, we adopt the implicit Euler discretization scheme for
the numerical simulation of \eqref{eq_growth}, yielding
\begin{multline*}
\frac{R_i^2(t+\Delta t_g)-R_i^2(t)}{2\Delta t_g} \\ 
= \frac{(S_i^*-1) - \frac{a(T_i^*)}{R_i(t+\Delta t_g)} 
        + \frac{b(M_i(t))}
       {R_i^3(t+\Delta t_g)}}{F_k(T_i^*) + F_d(T_i^*)},
\end{multline*}
where $S_i^*$ and $T_i^*$ are the ambient saturation ratio and the temperature at the
super-droplet position $\mathbf x_i(t)$, which are evaluated by the
fluid variables $q_v(\mathbf x_{lmn},t)$, $\theta(\mathbf x_{lmn},t)$,
and $\rho(\mathbf x_{lmn},t)$, applying interpolation.  It may be not
possible to determine analytically the next step value $R_i(t+\Delta
t_g)$ from the current value $R_i(t)$. Hence, we adopt the Newton-Raphson
scheme to evaluate $R_i(t+\Delta t_g)$. After evaluating $R_i(t+\Delta
t_g)$ for all the super-droplets $i$, we will update the mixing ratio
of vapor and potential temperature to $q_v(\mathbf x_{lmn},t+\Delta
t_g)$ and $\theta(\mathbf x_{lmn},t+\Delta t_g)$. Please refer to
Sec.~\ref{5.2} for the detail.

\subsubsection{Monte Carlo scheme for the coalescence of super-droplets}
\label{5.1.3}
Let $\Delta t_c$ be the time step to simulate this process.
Let us introduce a grid which covers the entire real-space for our
simulation. We can choose any kind of grid - which may be unstructured -
but the volume of each cell must be small enough that the
super-droplets are well-mixed inside the cell and the coalescence
occurs according to the probability \eqref{prob_super} in each
cell. Let us make a list of super-droplets in a certain cell at time
$t$: $I:=\{i_1, i_2, \dots , i_{n_s}\}$, where $n_s$ is the total number
of super-droplets in the cell.  Then, by definition, each pair of
super-droplets $(j,k)\in I^2$, $j\neq k$, coalesces with 
probability $P_{jk}^{(s)}$ within the short time interval $(t,t+\Delta
t_c)$.  Note that $P_{jk}^{(s)}\ll1$ if $\Delta t_c$ is small. We
have to examine all the possible pairs $(j,k)$ to
know the random realization at the time $t+\Delta t_c$.  However, this
yields computational cost $O(n_s^2)$ and this is not efficient.
Instead, we propose a novel Monte Carlo scheme which leads us to
$O(n_s)$ cost in computation.

Let
$L:=\{(j_{1},k_{1}),(j_{2},k_{2}),\dots,(j_{[n_s/2]},k_{[n_s/2]})\}$
be a list of $[n_s/2]$ randomly generated, non-overlapping
pairs.  Non-overlapping means that no super-droplet belongs to more
than one pair.  This list can be made by generating a random
permutation of $I$ and make pairs from the front, which cost $O(n_s)$
in computation (see Appendix~\ref{app.A}). By examining only these $[n_s/2]$
pairs instead of the whole possible $n_s(n_s-1)/2$ pairs, the cost
will be $O(n_s)$. In compensation for this simplification, the
coalescence probability is divided by the decreasing ratio of pair
number, and the corrected, scaled up probability for the $\alpha$th
pair is obtained as
\begin{equation*}
p_\alpha:=P_{j_\alpha k_\alpha}^{(s)}\frac{n_s(n_s-1)}{2}\Bigm/\left[\frac{n_s}{2}\right],
\;\;\; \alpha=1, 2, \dots, \left[\frac{n_s}{2}\right].
\end{equation*}
This may be justified if the following relation holds good, 
\begin{equation*}
\begin{split}
\sum_{\alpha =1}^{[n_s/2]}\min(\xi_{j_\alpha },\xi_{k_\alpha })p_\alpha  &\simeq
\frac{1}{2}\sumsum_{j,k,j\neq k}\min(\xi_j,\xi_k)P^{(s)}_{jk}\\
&=\frac{1}{2}\sumsum_{j,k,j\neq k}\xi_j\xi_kP_{jk},
\end{split}
\end{equation*}
which represent the consistency of the expectation number of coalesced
droplet pairs in this cell.

Based on the above idea, below is the complete procedure of our Monte
Carlo scheme to obtain one of the stochastic realizations in a certain
cell at time $t+\Delta t_c$ from the sate of super-droplets at time $t$.
\begin{enumerate}
\item Make the super-droplet list at
time $t$ in this certain cell: $I=\{i_1, i_2, \dots , i_{n_s}\}$.
\item Make the $[n_s/2]$ candidate-pairs
$L=\{(j_{1},k_{1}),(j_{2},k_{2}),\dots,(j_{[n_s/2]},k_{[n_s/2]})\}$
from the random permutation of $I$ (see Appendix~\ref{app.A} for detail).
\item For each pair of super-droplets $(j_\alpha,k_\alpha)\in L$,
generate a uniform random number $\phi_\alpha\in(0,1)$. Then,
evaluate
\begin{equation*}
  \gamma_\alpha:=
\begin{cases}
    [p_\alpha ]+1   & \text{if $\phi_\alpha <p_\alpha -[p_\alpha ]$}\\
    [p_\alpha ]     & \text{if $\phi_\alpha \ge p_\alpha -[p_\alpha ]$}
\end{cases}
\end{equation*}
\item If $\gamma_\alpha =0$, the $\alpha $th pair $(j_\alpha, k_\alpha)$ is updated to $t+\Delta t_c$
without changing their state.
\item If $\gamma_\alpha\neq0$, choose $\xi_{j_\alpha }\ge\xi_{k_\alpha }$ without losing
  generality and evaluate $\tilde\gamma_\alpha :=\min(\gamma_\alpha,[\xi_{j_\alpha }/\xi_{k_\alpha }])$.
    \begin{enumerate}
    \item if $\xi_{j_\alpha } -  \tilde\gamma_\alpha\xi_{k_\alpha }>0$,
      \begin{gather*}
	\xi'_{j_\alpha}=\xi_{j_\alpha}-\tilde\gamma_\alpha \xi_{k_\alpha },\quad\xi'_{k_\alpha }=\xi_{k_\alpha },\\
	R'_{j_\alpha}=R_{j_\alpha},\quad R'_{k_\alpha }=(\tilde\gamma_\alpha R_{j_\alpha}^3+ R_{k_\alpha }^3)^{1/3},\\
	M'_{j_\alpha}=M_{j_\alpha},\quad  M'_{k_\alpha}=(\tilde\gamma_\alpha M_{j_\alpha}+M_{k_\alpha }),\\
	\mathbf x'_{j_\alpha}=\mathbf x_{j_\alpha},\quad\mathbf x'_{k_\alpha }=\mathbf x_{k_\alpha },
      \end{gather*}
	   \item If $\xi_{j_\alpha } - \tilde\gamma_\alpha  \xi_{k_\alpha }=0$, 
i.e., $\tilde\gamma_\alpha =\xi_{j_\alpha }/\xi_{k_\alpha }=[\xi_{j_\alpha }/\xi_{k_\alpha }]\le\gamma_\alpha$,
\begin{gather*}
\xi'_{j_\alpha}=[\xi_{k_\alpha }/2],\quad \xi'_{k_\alpha }=\xi_{k_\alpha }-[\xi_{k_\alpha }/2],\\
R'_{j_\alpha}=R'_{k_\alpha }=\left(\tilde\gamma_\alpha  R_{j_\alpha}^3+ R_{k_\alpha }^3\right)^{1/3},\\
M'_{j_\alpha}=M'_{k_\alpha }=\left( \tilde\gamma_\alpha  M_{j_\alpha}+ M_{k_\alpha }\right),\\
\mathbf x'_{j_\alpha}=\mathbf x_{j_\alpha},\quad\mathbf x'_{k_\alpha }=\mathbf x_{k_\alpha }.
\end{gather*}	     
If $\xi'_{j_\alpha}=0$, the super-droplet $j_\alpha$ is removed out of
the system.
\end{enumerate}
\end{enumerate}

Our Monte Carlo scheme numerically solves the stochastic coalescence
process of super-droplets defined in Sec.~\ref{4.1.4}.  We reduced the
candidate pairs using a random permutation technique to achieve the
computational cost $O(n_s)$.  Further extension are introduced in our
Monte Carlo scheme to deal with {\it multiple coalescence}.  The
integer $\gamma_\alpha$ represents how many times the pair
$(j_\alpha,k_\alpha)$ will coalesce, and $\tilde\gamma_\alpha$ is the
restricted value of $\gamma_\alpha$ by its maximum value
$[\xi_{j_\alpha }/\xi_{k_\alpha }]$.  Primarily, $p_\alpha$ must be
smaller than $1$ because $p_\alpha$ represents a probability, thus
$\gamma_\alpha$ should be either $0$ or $1$. However, if we take
$\Delta t_c$ too much larger, $p_\alpha$ may become larger than
$1$. Though the situation $\gamma_\alpha>1$ is not consistent with the
fundamental premise of our theory, we formulate our Monte Carlo scheme
to cope with this situation so that SDM robustly well-approximates the
primitive model irrespective of the choice of $\Delta t_c$.

Alternatively, the most rigorous selection criteria of the
simulation time step for the coalescence process $\Delta t_c$ is that
$p_\alpha\ll1$ holds good for all the candidate pairs $\alpha$.  Let
us estimate $\Delta t_c$ very roughly.  The typical number density and
radius of small cloud droplets are $10^9~\mathrm{m}^{-3}$ and
$R=10~\mu \mathrm{m}$.  The typical terminal velocity of a rain
droplet is $1~\mathrm{m~s^{-1}}$.  Then, we may roughly evaluate
$p_\alpha$ as follows.
\begin{equation*}
\begin{split}
  p_\alpha &\simeq n_s P_{j_\alpha k_\alpha}^{(s)}\\
  &= \frac{n_s\max{(\xi_{j_\alpha},\xi_{k_\alpha})}}{\Delta V}E\pi(R_{j_\alpha}+R_{k_\alpha})^2\lvert\mathbf v_{j_\alpha}-\mathbf v_{k_\alpha}\rvert\Delta t_c\\
  &\simeq 10^9\text{m}^{-3}\cdot 1\cdot \pi (2\times10^{-5}\text{m})^2\cdot 1~\mathrm{m~s^{-1}}\cdot\Delta t_c\\
  &= (0.4\pi~\mathrm{s^{-1}})\Delta t_c.
\end{split}
\end{equation*}
Because $p_\alpha<1$ should be satisfied, we can estimate that $\Delta
t_c< 1/(0.4\pi~\mathrm{s^{-1}})\simeq0.8~\mathrm{s}$.  It is worth
noticing here, that our estimation is independent of the number of
super-droplets $n_s$ and the volume of the coalescence cell $\Delta
V$.

\subsubsection{Validation of our Monte Carlo scheme}
\label{5.1.4}
In this section, we validate our Monte Carlo scheme for the stochastic
coalescence process by confirming that the numerical results agree
well with the solution of a SCE.

The super-droplets in this section undergo
only the coalescence process, which will be simulated by our Monte
Carlo scheme. The super-droplets have velocities, but do not move out
of the coalescence cell, nor do they undergo
condensation/evaporation. The equivalent radius of water $R_i$ is the
only attribute.
The corresponding SCE for this extreme case can be derived as
\begin{equation}
\label{eq.13}
\begin{split}
\frac{\partial n(X,t)}{\partial t} = \;&\frac{1}{2} \int_{0}^{X} dX' n(X')n(X'')K(X',X'')\\
&- n(X)\int_{0}^{\infty} dX' n(X')K(X,X'),
\end{split}
\end{equation}
where $n(X,t)$ is the number density of droplets, 
$X=(4\pi/3)R^3$ is the water volume of droplet,
$K$ is the coalescence kernel defined in \eqref{prob}, 
and $X'':=X-X'$.

To compare the simulation result, we have to reconstruct the number
density distribution $n(X,t)$ from the data of super-droplets
$\{(\xi_i,R_i)\}$. There are several ways to do this, but we adopt
the kernel density estimate method with Gaussian kernel 
\citep{Terrell1992}. Some brief discussions on the application
of the kernel density estimate method to SDM is also developed in
Sec.~\ref{7.2}. It is convenient to plot the results by the mass density
function over $\ln R$, which is defined by $g(\ln R)d\ln
R:=\rho_\mathrm{liq}Xn(X)dX$.  The corresponding estimator function
$\Tilde g(\ln R)$ becomes,
\begin{equation*}
\begin{split}
\Tilde g(\ln R)&:=\frac{1}{\Delta V}\sum_{i=1}^{N_s}\xi_i m_i W_\sigma(\ln R-\ln R_i),\\
W_\sigma(Y)&:=\frac{1}{\sqrt{2\pi}\sigma}\exp{\left(-Y^2/2\sigma^2\right)}.\\
\end{split}
\end{equation*}
Here, $\Delta V$ is the volume of the coalescence cell, and
$\sigma=\sigma_0 N_s^{-1/5}$ with some constant
$\sigma_0$. Theoretically, the most efficient choice of $\sigma_0$ is
estimated as $\sigma_0=(2\sqrt{\pi}\Delta V^2\int g''^2d\ln R
/M_\mathrm{tot}N_r)^{-1/5}$, where $M_\mathrm{tot}$ is the total mass of liquid
water. Because $g$ itself is the function under estimation, we have to
estimate also $\sigma_0$ from the data of super-droplets to find the
maximum likelihood estimator function $\tilde g$, but in this paper we
simply choose $\sigma_0$ empirically.

\begin{figure}
\centering
\includegraphics[width=8.5cm]{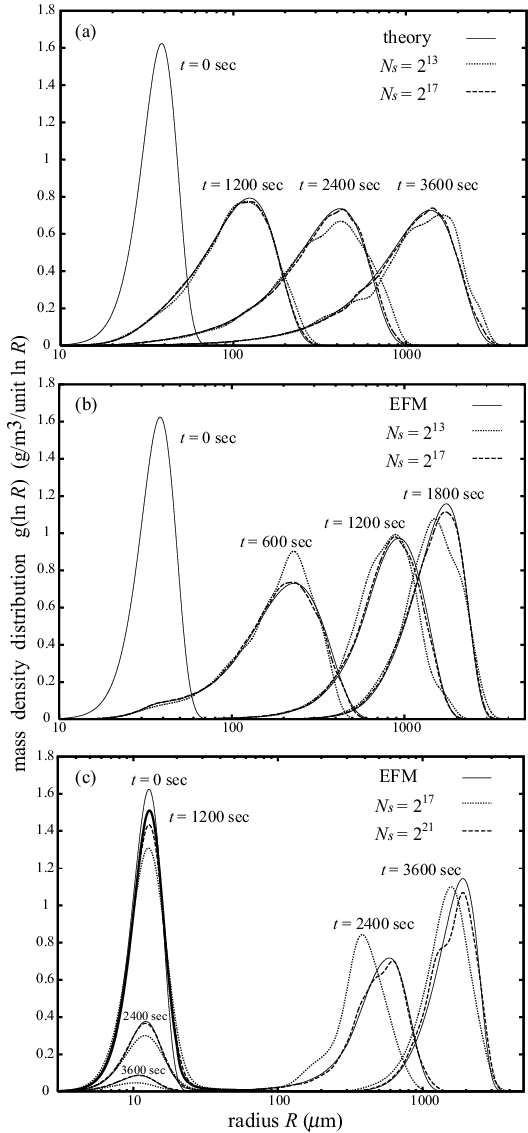}
\caption{Time evolution of the mass density distribution $g(\ln R,t)$,
which is numerically obtained by the Monte Carlo scheme of SDM. (a):
The case of Golovin's kernel.  The solid line represents the analytic
solution of the SCE~\eqref{eq.13}. (b) and (c): The case of
hydrodynamic kernel. The solid line represents the approximate
solution of the SCE~\eqref{eq.13}, which is numerically obtained by
EFM. In each case, we can see that the result of SDM agrees fairly
well with solution of the SCE~\eqref{eq.13} when the number of
super-droplets $N_s$ is sufficiently large. $\Delta
t_c=1.0~\mathrm{s}$ for (a) and (b), and $\Delta t_c=0.1~\mathrm{s}$
for (c).  }
\label{fig.golovin} 
\end{figure}

We have to give an explicit form of the coalescence kernel
$K(R_j,R_k)$ in \eqref{prob} to determine the dynamics.  
For Golovin's kernel $K(R_j,R_k)=b(X_j+X_k)$, we have the
analytic solution of the SCE~\eqref{eq.13} 
\citep{Golovin1963} though this choice of coalescence kernel is not
realistic.  The result of our comparison simulation for Golovin's
kernel is shown in Fig.~\ref{fig.golovin}a.  We set
$b=1.5\times10^3~\mathrm{s^{-1}}$. We need only use one big coalescence cell, 
the volume of which is chosen as $\Delta V=10^6~\mathrm{m}^{3}$.  
The time step is fixed as $\Delta t_c=1.0~\mathrm{s}$.  
The initial number density of droplets is set to be
$n_0=2^{23}~\mathrm{m}^{-3}$. The initial size distribution of the droplets
follows an exponential distribution of the droplet volume $X_i$
which is determined by the probability density 
%\begin{equation}
$
p(X_i)=(1/X_0)\exp{\left(-X_i/X_0\right)}$,
$X_0=(4\pi/3)R_0^3$,
%R_0=30.53096257018643\times10^{-6}\text{m},
$R_0=30.531\times10^{-6}~\mathrm{m}
%\end{equation}
$.
This setting results in the total amount of liquid water $1.0~\mathrm{g~m^{-3}}$.
%The line at $t=0\mathrm{s}$ represents initial distribution of droplets.  
The initial number of super-droplet $N_s$ is changed as the
simulation parameter. The initial multiplicity $\xi_i$ is determined
by the equality $\xi_i = n_0\Delta V/N_s$. $\sigma_0$ is fixed as $\sigma_0=0.62$.

Figures~\ref{fig.golovin}b and 2c show the results for the case of
the so-called hydrodynamic kernel, which is a much more realistic
coalescence kernel, defined by \eqref{prob}
and \eqref{col_eff}.  The same coalescence efficiency $E(R_j, R_k)$ is
adopted as described in 
\cite{Seeszelberg1996} and \cite{Bott1998}: For small droplets the
dataset of \cite{Davis1972} and 
\cite{Jonas1972} is used, and for large droplets the dataset
of \cite{Hall1980} is used.  Because the analytic
solution is not available for hydrodynamic kernel, we generated the
reference solution numerically using the Exponential Flux Method (EFM)
developed by \cite{Bott1998, Bott2000}. A
logarithmically equidistant radius grid is used in EFM. We adopted
1000 bins ranging from 0.62~$\mu$m to 6.34~cm. The time step was set
to 0.1~s. These parameter settings provide a sufficiently accurate
numerical solution of the SCE~\eqref{eq.13}.  In
Fig.~\ref{fig.golovin}b, simulation settings are the same as in case (a)
except for the coalescence kernel.  In Fig.~\ref{fig.golovin}c we changed
the initial size distribution by reducing $R_0$ to one-third, i.e.,
$R_0=10.177\times10^{-6}~\mathrm{m}$, and increased the initial number
density of droplets as
$n_0=3^3\cdot2^{23}~\mathrm{m}^{-3}$. Consequently, the total amount
of liquid water $1.0~\mathrm{g~m^{-3}}$ is unchanged from the case
(b). $\Delta t_c=0.1~\mathrm{s}$ and $\sigma_0=1.5$ for the case (c).

In case (a) and (b), we can see that the result of SDM with
$N_s=2^{13}$ agrees fairly well with the solution of
SCE~\eqref{eq.13}, and is much improved when $N_s=2^{17}$.  However, case
(c) is not so good as the previous two cases. Even $N_s=2^{17}$ is not
good and we need $N_s=2^{21}$ for good agreement.  This situation
reveals a typical weak point of SDM. The initial size of the
droplets is comparatively small and coalescence hardly occurs.
We can see that the thick solid line at $t=1200~\mathrm{s}$ is not so
changed from the initial distribution. However, once a few big
droplets are created, then the coalescence process is abruptly
accelerated. Accordingly, for an accurate prediction, we have to
resolve the right tail of the distribution at the time
$t=1200~\mathrm{s}$.  However, the existence probability of super-droplets
is very low for this region and sampling error occurs if the number of
super-droplets $N_s$ is not sufficient.  As a result, we need many
more super-droplets for the case (c) compared to that of (a) and
(b). For practical applications to cloud formation, this fact
would not impose severe restrictions on SDM because the
condensation/evaporation process dominates the growth of small
droplets in such cases like (c).

Anyway, we have confirmed that SDM reproduces the solution of the
SCE~\eqref{eq.13} if the number of super-droplets $N_s$ is
sufficiently large.  These results support the validity of our Monte
Carlo scheme for the stochastic coalescence.

\subsection{How to simulate cloud dynamics}
\label{5.2}
Let $\Delta t_f$ be the time step to simulate this process.
Let us introduce a regular grid for the spatial discretization of the
fluid field variables, and let $\mathbf x_{lmn}$ be the position
coordinate of the grid point $(l,m,n)$. Then, we can adopt the
4th-order Runge-Kutta scheme for the time derivative, the second-order
central difference scheme for the spatial derivatives, with artificial
viscosity to stabilize the numerical instability, to simulate
\eqref{motion}-\eqref{qv}.

Here, the cloud microphysics processes exert an influence on the
cloud dynamics through the source terms \eqref{eq.src_super1} and
\eqref{eq.src_super2}.  While we calculate the momentum coupling
related to \eqref{eq.src_super1} in the fluid time step $\Delta t_f$, we
calculate the vapor and latent heat coupling related to
\eqref{eq.src_super2} in the condensation/evaporation time step
$\Delta t_g$, in the following manner.

Every time, before updating the fluid variables in the
$\Delta t_f$ time step, we evaluate the momentum coupling term
\eqref{eq.src_super1} from the current state of the super-droplets.
Here, \eqref{eq.src_super1} must be evaluated on the numerical grid point 
$\mathbf x_{lmn}$. Generally, by introducing an appropriately
chosen coarsening function $w(\mathbf x)$, it can be evaluated as
\[
\begin{split}
\rho_w(\mathbf x_{lmn},t)&=\int w(\mathbf x_{lmn}-\mathbf x) \rho_w(\mathbf x,t) d^3x\\
&=\sum_{i=1}^{N_s}\xi_im_i(t)w(\mathbf x_{lmn}-\mathbf x_i(t)).
\end{split}
\]
Here $w$ satisfies $\int w(\mathbf x)d^3x=1$, and would be a piecewise
linear, localized function with length-scale comparable to the grid
width. Our simplest choice is to share each super-droplet by the 8 adjacent grid
points at the corner of the cell, then $w(\mathbf x)=1/8\Delta V$ inside
the cube with volume $8\Delta V$, and $w(\mathbf x)=0$ outside the cube.
Then, using this $\rho_w(\mathbf x_{lmn},t)$, we perform the
4th-order Runge-Kutta scheme to update the fluid variables to the time 
$t+\Delta t_f$.

On the other hand, we calculate the increase and decrease of vapor and potential
temperature through \eqref{eq.src_super2} after we have
calculated the condensation/evaporation of super-droplets in the time
step $\Delta t_g$ (see also Sec.~\ref{5.1.2}).  Similarly to
$\rho_w(\mathbf x_{lmn},t)$, $S_v$ on the grid point $\mathbf x_{lmn}$
is evaluated as
\begin{multline*}
S_v(\mathbf x_{lmn},t)=\frac{-1}{\rho(\mathbf x_{lmn},t)}\\
\times\sum_{i=1}^{N_s}\xi_i\frac{m_i(t+\Delta t_g)-m_i(t)}{\Delta t_g}w(\mathbf x_{lmn}-\mathbf x_i(t)).
\end{multline*}
Then, the mixing ration of vapor and the potential temperature are updated as
\begin{gather*}
\theta(\mathbf x_{lmn},t+\Delta t_g)=\theta(\mathbf x_{lmn},t)-\Delta t_g\frac{LS_v(\mathbf x_{lmn},t)}{c_p\Pi(\mathbf x_{lmn},t)},\\
q_v(\mathbf x_{lmn},t+\Delta t_g)=q_v(\mathbf x_{lmn},t)+\Delta t_gS_v(\mathbf x_{lmn},t).
\end{gather*}

The way how to simulate the primitive model using the SDM has now been
obtained completely.

%%%%%%%%%%%%%%%%%%%%%%%%%%%%%%%%%%%%%%%%%%%%%%%
% Demonstration
\section{Proof of concept}
\label{6}
A simple test of a shallow maritime cumulus formation initiated by a warm bubble
is presented in this section to evaluate the behavior of our model.

The simulation domain is 2-dimensional ($x$-$z$), 12.8~km in the horizontal
direction and 5.12~km in the vertical direction.  Initially, a humid but
not saturated atmosphere is stratified, which is almost absolutely
unstable under 2~km in altitude. There is a temperature inversion layer
above 2~km.  The equations below determine the initial base sounding of the
atmosphere.
\begin{gather*}
\mathbf U(\mathbf x,t=0)=0,\\
T(x,z=0,t=0)=305.0~\mathrm{K},\\
P(x,z=0,t=0)=101325~\mathrm{Pa},\\
\frac{\partial T}{\partial z} = \begin{cases}
  -9.5~\mathrm{K~km^{-1}} & \text{if $z<2.0~\mathrm{km}$},\\
  +3.0~\mathrm{K~km^{-1}} & \text{if $z\ge2.0~\mathrm{km}$},
                                \end{cases}\\
q_v(\mathbf x,t=0)=0.022\exp{\left[-\left\{\frac{(z+200.0~\mathrm{m})}{2000.0~\mathrm{m}}\right\}^2\right]}.
\end{gather*}
The upper and lower boundaries are fixed to their initial values and the
horizontal boundary is periodic.  A warm bubble is inserted at  the
center of the horizontal axis according to the equation below.
\begin{multline*}
\theta(\mathbf x,t=0)=\theta_b(\mathbf x,t=0)+1.0~\mathrm{K}\\
\times\exp{\left[-\left\{\frac{(z-500.0~\mathrm{m})}{400.0~\mathrm{m}}\right\}^2
-\left\{\frac{(x-6.4~\mathrm{km})}{1200.0~\mathrm{m}}\right\}^2\right]},
\end{multline*}
where $\theta_b(\mathbf x,t=0)$ is the base state of potential temperature.

We simulate the cloud dynamics according to the non-hydrostatic model
introduced in Sec.~\ref{2.2} with the numerical schemes proposed in
Sec.~\ref{5.2}. We use a regular grid with the mesh size $\Delta
x=3.125~\mathrm{m}$ and $\Delta z=4.0~\mathrm{m}$. The time step is
$\Delta t_f= 0.0125~\mathrm{s}$.  We set the artificial viscosity
$\nu=\kappa=1.5625~\mathrm{m^2~s^{-1}}$.

Initially, NaCl aerosols are uniformly distributed in space with number
density $1.0\times10^7~\mathrm{m}^{-3}$.  The solute mass $M_i(t=0)$
has an exponential distribution given by the probability density
$p(M_i)=(1/M_0)\exp{(-M_i/M_0)}$, $M_0=1.0\times10^{-16}~\mathrm{g}$.
Because it is unsaturated at the beginning, the growth
equation~\eqref{eq_growth} has a stable and steady solution, which is
adopted as the initial value of the equivalent radius of water
$R_i(t=0)$. The coalescence efficiency used in Sec.~\ref{5.1.4} is also
adopted to this simulation.

For the numerical simulation of the microphysical processes, we also
use the numerical schemes proposed in Sec.~\ref{5}.  The same grid as that
of the non-hydrostatic model is adopted for our Monte Carlo scheme.
Initially, the super-droplets are uniformly distributed in space and
the number density is chosen as $64$ per cell.  We consider that we
have a single cell in the $y$-direction of dimension $100~\mathrm{m}$, then the
number density of super-droplets is equivalent to
$64/\mathrm{cell}=5.12\times10^{-2}~\mathrm{m}^{-3}$.  
For the super-droplets, we impose a common initial multiplicity, 
which value  is determined by
$\xi_i(t=0)=~\text{INT}(1.0\times10^7/5.12\times10^{-2})=195312500$.
The time steps are $\Delta t_m = \Delta t_g = \Delta t_c= 1.0~\mathrm{s}$.

\begin{figure*}
\centering
\includegraphics[width=\textwidth]{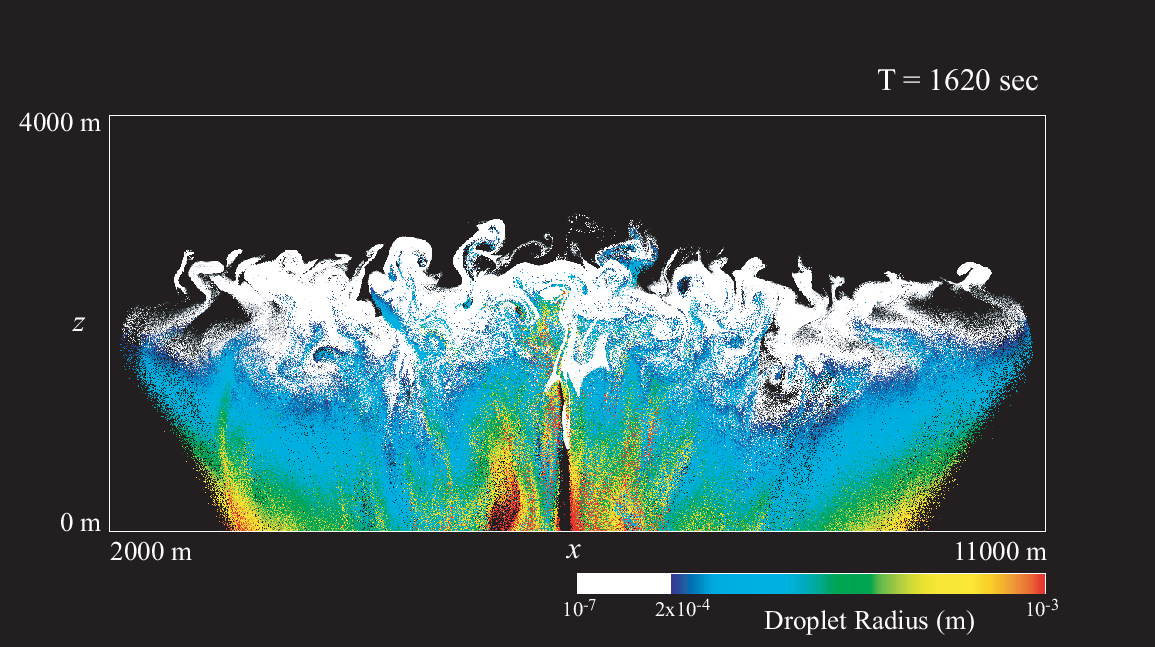}
\caption{Shallow maritime cumulus formation initiated by a warm bubble
simulated by SDM. This is a snapshot at time $1620$~s. The
super-droplets are plotted with the color and the alpha transparency
which are determined by their radius $R_i$ and multiplicity
$\xi_i$. The color map is indicated in the figure and the alpha value
is determined by
$\alpha_i=1.0\times10^{-5}(R_i/10^{-3}~\mathrm{m})^2~\xi_i$, which is
proportional to $R_i^2\xi_i$. We can see that the turbulent
like structures inside the cloud is resolved in our simulation.
Note that
super-droplets are also in the black region, but they are not
depicted in this figure because they are regarded as aerosols.
\label{fig.cloud}
}
\end{figure*}

Cloud started to form at about $8$~min and began to rain at about
$20$~min.  After raining for half an hour, the cloud remained for a
while but finally disappeared at about $90$~min. The amount of
precipitation is about $1$~mm in total. Figure~\ref{fig.cloud} is a
snapshot at time $1620$~s. We plot the super-droplets with the
color and the alpha transparency which are determined by their radius
$R_i$ and multiplicity $\xi_i$. The color map is indicated in the
figure and the alpha value is determined by
$\alpha_i=1.0\times10^{-5}(R_i/10^{-3}~\mathrm{m})^2~\xi_i$, which is
proportional to $R_i^2\xi_i$.  We can see that the turbulent-like
structures inside the cloud are resolved in our simulation. Note that
there are super-droplets also in the black region, but it is not
depicted in this figure because the water amount contained in the
super-droplets is very low; these super-droplets are aerosols.

To perform an accurate simulation, the number of super-droplets $N_s$
must be sufficiently large. However, it is not so easy to determine the
optimal number of super-droplets to be used.  A lower bound is to use
about $100$ super-droplets per fluid cell, otherwise
the source terms $\rho_w(\mathrm x_{lmn},t)$ and $S_v(\mathrm
x_{lmn},t)$ would be not smooth enough in space compared to the fluid
variables.  An upper bound is to use about $100,000$ 
super-droplets for each coalescence cell if the number of attributes
$d=1$. Then the droplet size distribution in each coalescence
cell can be fully resolved, as we have already discussed in
Sec.~\ref{5.1.4}.  In our simulation, we used only $64$
super-droplets for each coalescence cell, nevertheless we found that
the overall behavior, such as the cloud shape and the lifetime,
fairly well converged. Further analyses have to be carried out
to confirm the convergence of our simulation by checking the detailed
structures of the cloud, such as the droplet size distribution and the
aerosol size distribution.

\section{Future Direction}
\label{7}
\subsection{Extension of Super-Droplet Method}
\label{7.1}
The model we developed in Sec.~\ref{4} and \ref{5} was only for warm rain with only one
soluble substance as the aerosol. However, it is very conceivable that we
can extend SDM to incorporate various cloud microphysical processes,
such as, several sorts of soluble/insoluble aerosols, their chemical
reactions, several types of ice crystals, electrification, and the
breakup of droplets. Here, the possible extensions of the method is briefly discussed.

For this purpose, let us generalize the definition of attributes as
$\mathbf a(t) =
(a^{(1)}, a^{(2)},\dots,a^{(d)})$, where $d$ is the number of independent
attributes. Then, we have to construct the time evolution law of attributes that determines 
the individual dynamics of super-droplets, which can be written as:
\begin{equation}
\label{eq.a}
  \frac{d\mathbf a_i}{dt}=\mathbf f(\mathbf a_i,\mathbf A(\mathbf x_i)),\quad i=1,2,\dots,N_s(t),
\end{equation}
where $\mathbf A(\mathbf x_i)$ represents the field variable which
characterizes the state of the ambient atmosphere.  In the
case of the primitive model, eq.~\eqref{eq.a} consists of two
equations: the growth equation~\eqref{eq_growth} and the time
evolution of solute mass $dM_i/dt=0$. 

The stochastic coalescence process have to be redefined for our new super-droplets.
At first, we have to determine what kind of droplet $\mathbf a'$ will
be created after the coalescence of droplets $\mathbf a_j$ and $\mathbf
a_k$.  
Then, the
coalescence of super-droplets will be determined similarly to
\eqref{eq.15}-\eqref{eq.22}. It is not the case in
\eqref{eq.18} and \eqref{eq.22}, but we may also change the
position of the super-droplets after the coalescence if necessary. Similar
to \eqref{prob}, the coalescence probability for droplets
could be evaluated in the form
\begin{equation}
\label{eq.prob.gen}
\begin{split}
P_{jk} 
=& C(\mathbf a_j, \mathbf a_k)
\frac{\Delta t_c}{\Delta V} | \mathbf v_j - \mathbf v_k|\\ 
=&\mbox{probability that droplet $j$ and $k$ } \\ 
 &\mbox{inside a small region $\Delta V$ will coalesce }\\
 &\mbox{in a short time interval $(t,t+\Delta t_c)$,}
\end{split}
\end{equation}
which in general is a function of the attributes $\mathbf a_j$ and $\mathbf
a_k$.  Then the coalescence probability of super-droplets will be
determined by \eqref{prob_super}.

Incorporating the breakup of droplets to the SDM framework is a
remaining problem to be solved. Such a formulation would be desirable
that the number of super-droplets do not change after a breakup event,
but this is our future task.

In various research areas, many types of particle-based simulation
schemes have been developed recently.  SDM can be regarded as a
sort of Direct Simulation Monte Carlo (DSMC) method, which was
initially proposed to simulate the Boltzmann equation for predicting
rarefied gas flows \citep{Bird}.  In particular, SDM has
similarity to the Extended version of the No-Time-Counter (ENTC)
method, which was developed by \cite{Schmidt2000} for the
simulation of spray flows. Though the basic idea
of using a computational particle with varying multiplicity is common
in both SDM and ENTC, the Monte Carlo procedures to simulate the
stochastic coalescence process are different:  SDM is using $[n_s/2]$ randomly
generated, non-overlapping candidate pairs, and
allows multiple coalescence for each pair.  On the other hand, the ENTC
method chooses the candidate pairs randomly from the set of all the
possible pairs. The number of candidate-pairs to be sampled varies
proportionally to the maximum coalescence probability in the cell.
Both the SDM and the ENTC methods result in the computational cost
$O(N_s)$. Presumably, the ENTC method is also applicable to cloud
simulations and SDM is applicable to spray simulations, but further
verifications are still necessary to compare their capability.

Independently to the work by the present authors, there exist several
ongoing studies to construct novel cloud microphysics models employing
similar ideas as SDM. \cite{Andrejczuk} have developed a
particle-based cloud microphysics model, in which they have implemented the
ice phase processes and some aerosol chemistry processes. In their
paper, the stochastic coalescence process was not incorporated yet,
but they are currently developing an original deterministic scheme to
calculate the stochastic coalescence process (2008, personal
communication).  On the other hand, a work employing a probabilistic
approach to accelerate the spectral (bin) method is currently in
progress by Sato et al. (2008, personal communication).

\subsection{Computational cost and accuracy}
\label{7.2}
The possible extension of the SDM has been discussed in the previous
section. We can see that once the fundamental equations of the cloud
microphysical processes, \eqref{eq.a} and \eqref{eq.prob.gen}, are
given, we can perform an accurate simulation using the SDM. 
However, this statement is also true for the spectral (bin) method.
Then, which method is computationally more efficient?

To answer this question, we have carried out a simple theoretical
analysis to compare the computational cost of the SDM and the
spectral (bin) method, which is presented in Appendix~\ref{app.B}.
Here, the costs are evaluated by the amount of the operation count and
memory necessary to reproduce the number density distribution of droplets within
a certain margin of error.

The assertions must still be verified, but our estimation suggests
that the computational cost of SDM could be less demanding than the
spectral (bin) method when the number of attributes $d$ is larger than
some critical value $d_c$.  In the case when the attribute-spatial
discretization error of the spectral (bin) method is in the range
$1\sim2$, then the critical value $d_c$ is estimated to be in the
range $2\sim4$.

Finally, we would like to present some supplemental data on the actual CPU behavior of SDM.
In Sec.~\ref{5.1.4}, we tested our Monte Carlo scheme and compared the results with the spectral (bin) method.
In case (b), 
the FLoating point OPeration (FLOP) count was $2.7~\mathrm{G}$ and the memory used was $192~\mathrm{MB}$ for $N_s=2^{13}$. 
For $N_s=2^{17}$, the FLOP count was $4.2\times10~\mathrm{G}$ and the memory used was $208~\mathrm{MB}$.
We produced the reference solution by EFM, the spectral (bin) method developed by Bott,
using $N_b=1000$ bins. 
For this simulation, 
the FLOP count was $6.5\times10^2~\mathrm{G}$ and the memory used was $80~\mathrm{MB}$.
Practically, $N_b=100$ would be sufficient and in this setting
the FLOP count is $7.5~\mathrm{G}$ and the memory used is $48~\mathrm{MB}$,
and we need $13~\mathrm{s}$ in wall-clock time to complete this simulation if we use one
Intel Pentium 4 CPU $3.20~\mathrm{GHz}$.

In Sec.~\ref{6} we performed a 2D simulation of a shallow maritime cumulus
formation.  In this simulation, the FLOP count was 
$9.3\times10^6~\mathrm{G}$ and the memory used was $65~\mathrm{GB}$.
We used $32$ nodes of the Earth Simulator, using $11$ hours of wall-clock
time to complete $2.5$ hours of simulated time, which means that the
performance of our calculation was $2.3\times10^2~\mathrm{GFLOPS}$ (Giga
FLoating point Operations Per Second). This is $11~\%$ of the peak performance.

%%%%%%%%%%%%%%%%%%%%%%%%%%%%%%%%%%%%%%%%%%%%%%%
\section{Summary and concluding remarks}
\label{8}
In this paper, we proposed the SDM as a new approach for the accurate
simulation of clouds and precipitation.

The primitive model was introduced at the beginning, which is a
detailed microphysics-dynamics coupled warm-cloud model. In this paper
the primitive model was regarded as the master framework for
describing the behavior of warm clouds.  We saw how the traditional
methods can simulate the primitive model and what are the problems
involved.

Then, we applied the SDM to the primitive model. First, introducing the
notion of super-droplets, we derived the basic equations of SDM and
discussed the ideas of SDM. Next, a numerical implementation
scheme was proposed for each process.  In particular, a novel Monte
Carlo scheme for the stochastic coalescence process was developed and
validated by comparing the numerical results with the solutions of
SCE.

To evaluate the behavior of our model, we performed a simple testing simulation
of a shallow maritime cumulus formation initiated
by a warm bubble.  Brief discussions were carried out on the
possible extensions of the SDM to incorporate various other cloud microphysical
processes.  Our theoretical estimation suggested
that SDM becomes computationally more efficient than the spectral (bin)
method when the number of attributes $d$ becomes larger than some
critical value, which we estimate to be in the range $2\sim4$.

Though several extensions and validations are still necessary, we
expect that SDM would be more feasible for the modeling of complicated
cloud microphysical processes, and provide us a new approach to the
open problems in cloud science, such as cloud and aerosol
interactions, cloud-related radiative processes, and the mechanism
of thunderstorms and lightning.

\ack
The authors are grateful to A. Bott for offering his Fortran program,
F. Araki for fruitful discussions, and T. Sato for his continued
support.  S.S. would like to thank T. Miyoshi, Y. Kawamura, S. Koyama, S. Hirose,
H. Nakao, and R. Onishi for valuable comments and encouragement. This
work is one of the results of the Earth Simulator project
``Development of Multi-scale Coupled Simulation Algorithm.''

\begin{appendix}
%\section*{APPENDIX: How to make a random permutation}
\section{How to make a random permutation}
\label{app.A}
Let $a^{(n)}=(a^{(n)}_1,a^{(n)}_2,\dots,a^{(n)}_n)$ be a permutation
of $I_n:=(1,2,\dots,n)$. Let $S_n:=\{a^{(n)}\}$ be the set of all the
permutation of $I_n$.  Thus, $S_n$ has $n!$ number of elements.  Let
us define {\it the random permutation of $I_n$} by such a permutation
$a^{(n)}$ that all the $a^{(n)}\in S_n$ occurs with the same
probability, i.e., $P(a^{(n)})=1/n!$, $\forall a^{(n)}\in S_n$.

We have two propositions. 1) $a^{(1)}=(1)$ is a random permutation of
$I_1$.  2) We can make a random permutation of $I_{n+1}$ from a random
permutation of $I_n$ by the following procedure: Let $a^{(n)}$ be a
random permutation of $I_n$ and choose $m$ from $1, 2, \dots, n+1$
randomly. Let $a^{(n+1)} = (a^{(n)},
n+1)=(a^{(n)}_1,\dots,a^{(n)}_n,n+1)$.  If $m\ne n+1$, exchange the
value at $m$ and $n+1$, i.e., $a^{(n+1)}_{n+1}=a^{(n)}_m$,
$a^{(n+1)}_{m}=n+1$. Then, $a^{(n+1)}$ is a random permutation of
$I_{n+1}$.

Proposition 1) is obvious and 2) is also easy to prove. For all
$a^{(n+1)}\in S_{n+1}$, we can uniquely determine $a^{(n)}\in S_n$ which
have the possibility to become $a^{(n+1)}$. Indeed, such $a^{(n)}$ is given
by
$a^{(n)}=(a^{(n+1)}_1,a^{(n+1)}_2,\dots,a^{(n+1)}_{m-1},a^{(n+1)}_{n+1},a^{(n+1)}_{m+1},\dots,a^{(n+1)}_n)$,
here $m$ satisfies $a^{(n+1)}_m=n+1$. If $m=n+1$,
$a^{(n)}_i=a^{(n+1)}_i$, $i=1,\dots,n$. Then, the probability that
$a^{(n+1)}$ occurs is evaluated by $P(a^{(n+1)})=P(a^{(n)})\times
1/(n+1)=1/(n+1)!$.

Consequently, we can make a random permutation of $I_n$ with $O(n)$
operations. 

\section{Estimation of the computational cost of the spectral (bin) method and the SDM}
\label{app.B}
First, let us estimate the computational cost and accuracy of the spectral
(bin) method.  We have extended SDM in Sec.~\ref{7.1}, and the
corresponding, more general form of SCE can be derived as follows.
\begin{gather}
\label{eq.SCE_general}
\frac{\partial n(\mathbf a,\mathbf x, t)}{\partial t} 
+\nabla_x\cdot\left\{\mathbf vn\right\}
+\nabla_a\cdot\left\{\mathbf fn\right\} 
= \left(\frac{\partial n}{\partial t}\right)_c,\\
\begin{split}
\left(\frac{\partial n}{\partial t}\right)_c
:=~&\frac{1}{2}\int d^da'
 n(\mathbf a')n(\mathbf a'')K(\mathbf a',\mathbf a'')\\\notag
&- n(\mathbf a)\int d^da' n(\mathbf a')K(\mathbf a,\mathbf a').
\end{split}
\end{gather} 
where $n(\mathbf a,\mathbf x,t)$ is the number density distribution, $\mathbf
f$ is defined in \eqref{eq.a}, and $\mathbf a''$ is such attribute that will become $\mathbf a$ if it coalesces with $\mathbf a'$.  The term $(\partial n/\partial t)_c$
corresponds to the ensemble-averaged dynamics of the stochastic
coalescence process, and this term is $d$-multiple integral in general
because we have to survey all over the $d$-dimensional attribute-space
to evaluate the time evolution of the droplet number density at
$\mathbf a$. Obviously, this term is computationally most expensive.
Therefore, to simplify the discussions, we neglect the advection terms
in the l.h.s of \eqref{eq.SCE_general} and focus our attention on the
simplified equation $\partial n(\mathbf a, t)/\partial t=(\partial n/\partial t)_c$,
which corresponds to the extreme situation that only the stochastic
coalescence is taken into account as the microphysical process.

Let $N_b$ be the number of bins for each attribute, i.e.,
the number of grid points used for the discretization of $n(\mathbf a)$ per
attribute. Then, because $\partial n/\partial t=(\partial n/\partial t)_c$ is a 
$d$-multiple integro-differential equation, the {\it operation count} and {\it memory} required for the
computation of spectral (bin) method scales like ``$\text{\it operation} \sim N_b^{2d}$'' 
and ``$\text{\it memory} \sim N_b^{d}$.''

To evaluate the accuracy of the spectral (bin) method, let us introduce
the Integrated Squared Error (ISE),
\begin{equation}
\label{eq.ISE}
C=\int d^da\;\left\{n(\mathbf a)- n_b(\mathbf a)\right\}^2.
\end{equation}
Here, $n_b(\mathbf a)$ is the approximate solution generated by the
spectral (bin) method, and $C$ measure the difference of $n_b(\mathbf a)$
from the exact solution $n(\mathbf a)$.  If the discretization error of
the spectral (bin) method is $k$th order in attribute-space, $C$ scales like
$C\sim N_b^{-2k}$ by definition.

Combining the above mentioned results, we can estimate the scaling of
{\it operation count} and {\it memory} needed for the computation of
spectral (bin) method in terms of $C$ as follows,
\begin{equation}
\label{eq.41}
\begin{split}
\text{\it operation} &\sim N_b^{2d} \sim \left(\frac{1}{\sqrt{C}}\right)^{2d/k},\\
\text{\it memory} &\sim N_b^{d} \sim \left(\frac{1}{\sqrt{C}}\right)^{d/k}.
\end{split}
\end{equation}

Next, let us estimate the computational cost and accuracy of SDM.  To
compare the result with the spectral (bin) method, we only consider the
stochastic coalescence process, and estimate the number density
distribution $n(\mathbf a)$ from the super-droplets $\{(\xi_i,\mathbf a_i)\mid
i=1,2,\dots,N_s\}$.  For the estimation, we use the kernel density
estimation method, which was originally developed to estimate the
generating probability distribution from its random sample \citep{Terrell1992}.

We use the density estimator function $\Tilde n(\mathbf a)$ with Gaussian
kernel $W_\sigma^{(d)}(\mathbf a)$, defined by
\begin{equation}
\label{eq.estimator}
\begin{split}
\Tilde n(\mathbf a)&:=\sum_{i=1}^{N_s}\xi_i W_\sigma^{(d)}(\mathbf a-\mathbf a_i),\\
W_\sigma^{(d)}(\mathbf a)&:=\frac{1}{(\sqrt{2\pi}\sigma)^d}\exp{\left\{-\mathbf a^2/2\sigma^2\right\}}.
\end{split}
\end{equation}
The error evaluation function corresponding to ISE~\eqref{eq.ISE} is the Mean Integrated
 Squared Error (MISE) defined by
\begin{equation}
\label{eq.MISE}
C(\sigma)=E\left[\int d^da\;\left\{n(\mathbf a)-\Tilde n(\mathbf a)\right\}^2 \right].
\end{equation}
Note here that $C(\sigma)$ is defined as an ensemble-averaged value
because each $\{(\xi_i,\mathbf a_i)\}$ is one of the random realizations
of the stochastic coalescence process.

Let $q(\xi,\mathbf a, t; N_s)$ be the number density of super-droplets,
i.e., $q(\xi,\mathbf a, t; N_s) \Delta^da$ is the expectation number of
super-droplets with multiplicity $\xi$ and attribute in the small
interval $(\mathbf a, \mathbf a+\Delta^da)$ at time $t$.  Obviously,
$q(\xi,\mathbf a, t; N_s)$ depends on the total number of super-droplets
$N_s$.  Let us assume that the following form of scaling law exists:
$q(\xi,\mathbf a, t; \alpha N_s)=\alpha^{k_1}q(\alpha^{k_2} \xi,\mathbf a, t; N_s)$, and
also assume that super-droplets can describe the expected dynamics of the
droplets irrespective of the choice of $N_s$, i.e.,
$\sum_{\xi=0}^{\infty}\xi q(\xi,\mathbf a,t;N_s)=n(\mathbf a,t)$.  Then, remembering the equality 
$\int d^da\sum_{\xi=0}^{\infty}q(\xi,\mathbf a,t;N_s)=N_s$,
we can determine the scaling exponents as $(k_1,k_2)=(2,1)$.  Based on this result, we
can derive the expectation value and variance of $\Tilde n(\mathbf a)$ as
\begin{gather*}
 E[\Tilde n(\mathbf a)]\simeq n(\mathbf a)  + \frac{\sigma^2}{2}\left\{\sum_j\frac{\partial^2 n(\mathbf a)}{\partial a_j^2}\right\},\\
 V[\Tilde n(\mathbf a)]\simeq\frac{N_r n(\mathbf a)}{N_s(2\sqrt{\pi}\sigma)^d}.
\end{gather*}
Substituting these two equations into \eqref{eq.MISE}, we can
determine the $\sigma^*$ that minimizes the MISE $C(\sigma)$, 
which yields the scaling,
\begin{equation*}
\sigma^*\sim N_s^{\frac{-1}{(d+4)}},\quad C(\sigma^*)\sim N_s^{\frac{-4}{(d+4)}}.
\end{equation*}
Thus, the {\it operation count} and {\it memory} needed for SDM scales like
\begin{equation}
  \label{eq.48}
  \begin{split}
    \text{\it operation} \sim N_s &\sim \left(\frac{1}{\sqrt{C(\sigma^*)}}\right)^{(d+4)/2},\\
    \text{\it memory} \sim N_s &\sim \left(\frac{1}{\sqrt{C(\sigma^*)}}\right)^{(d+4)/2}.
  \end{split}
\end{equation}

Now, we are ready to compare the computational efficiency of SDM and the spectral (bin) method by
comparing the exponents in \eqref{eq.41} and \eqref{eq.48}.
The result suggests that the operation count of SDM becomes
lower than the spectral (bin) method when the condition
\begin{equation*}
d> \frac{4k}{4-k}\quad\text{and}\quad k < 4,
\end{equation*}
is satisfied, and the memory of SDM becomes lower than the spectral (bin) method when
\begin{equation*}
d> \frac{4k}{2-k}\quad\text{and}\quad k < 2.
\end{equation*}

Thus, if we assume that the attribute-spatial discretization error of
the spectral (bin) method $k=1\sim2$, SDM becomes more efficient in
operation count when $d>2\sim4$, and becomes more efficient in memory when
$d>4\sim\infty$.  This result may be improved if we choose a different
kernel function $W(\mathbf a)$ for the density estimator $\Tilde n(\mathbf
a)$.

\end{appendix}

\bibliographystyle{wileyqj}
\bibliography{ref}

\end{document}